\documentclass[pra,twocolumn,superscriptaddress,showpacs]{revtex4}
\usepackage{amsmath,amssymb,amsfonts,bbm,graphicx}
\usepackage[dvips,usenames]{color}
\def\Tr{\hbox{Tr}}
\newcommand{\ketbra}[2]{\vert #1 \rangle \! \langle #2 \vert}
\newcommand{\braket}[2]{\langle #1 \vert #2 \rangle}
\newcommand{\ket}[1]{\vert #1 \rangle}
\begin{document}
\title{Enhancement of parameter estimation by Kerr interaction}
\author{Marco G. Genoni}\email{marco.genoni@fisica.unimi.it}
\affiliation{CNSIM, UdR Milano, I-20133 Milano, Italy}
\affiliation{Dipartimento di Fisica, Universit\`a degli Studi di Milano,
I-20133 Milano, Italy}
\author{Carmen Invernizzi}\email{carmen.invernizzi@unimi.it}
\affiliation{Dipartimento di Fisica, Universit\`a degli Studi di Milano,
I-20133 Milano, Italy}
\affiliation{CNSIM, UdR Milano, I-20133 Milano, Italia}
\author{Matteo G. A. Paris}\email{matteo.paris@fisica.unimi.it}
\affiliation{Dipartimento di Fisica, Universit\`a degli Studi di Milano,
I-20133 Milano, Italy}
\affiliation{CNSIM, UdR Milano, I-20133 Milano, Italy}
\affiliation{ISI Foundation, I-10133 Torino, Italy}
\date{\today}
\begin{abstract}
We address quantum estimation of displacement and squeezing parameters
by the class of probes made of Gaussian states undergoing Kerr
interaction. If we fix the overall energy available to the probe,
without posing any constraint on the available Gaussian squeezing, then
Gaussian squeezing represents the optimal resource for parameter
estimation.  On the other hand, in the more realistic case where the
amount of Gaussian squeezing is fixed,  or even absent, then Kerr
interaction turns out to be useful to improve estimation, especially for
probe states with large amplitude.  Our results indicate that precision
achievable with current technology Gaussian squeezing may be attained
and surpassed for realistic values of the Kerr coupling.
\end{abstract}
\maketitle
\section{Introduction}\label{s:intro}
Nonclassical states of light represent a resource for high-precision
measurements. They are generally produced in active optical media, which
couple one or more modes of the field through the nonlinear
susceptibility of the matter.  In particular, parametric processes in
second order $\chi^{(2)}$ media correspond to Gaussian operations and
are used to generate squeezing, hereafter {\em Gaussian} squeezing, and
entanglement.  Gaussian squeezing is the basic ingredient of quantum
enhanced interferometry \cite{bondurant,caves,yurke,
bandilla,hradil,paris,x02} and found several applications in quantum
metrology and communication \cite{tlp1,tlp2,tre03,sch05,slb05,sch07}.
In addition, Gaussian squeezing is the key resource to achieve precise
estimation of unitary \cite{Mon06,Gai09} and non unitary parameters
\cite{Mon07}. In turn, squeezed vacuum state has been addressed as a
universal optimal probe \cite{bm99,Mon07,Gai09} within the class of
Gaussian states. 
\par
On the other hand, the Kerr effect taking place in third-order nonlinear
$\chi^{(3)}$ media leads to a non Gaussian operation, and has been
suggested to realize quantum nondemolition measurements
\cite{Wal92,bra}, and to generate quantum superpositions
\cite{KerrYS,KK1,KK2} as well as squeezing \cite{Sun96} and entanglement
\cite{KerrLeuchs}.  A well known example of Kerr media are optical
fibers where, however, nonlinearities are very small and accompanied by
other unwanted effects. Larger Kerr nonlinearities have been observed
with  electro- magnetically induced transparency \cite{KerrEIT} and with
Bose Einstein condensates \cite{KerrBEC} and cold atoms  \cite{KerrCA}.
Recently, nonlinearities on 9 orders of magnitude higher than natural
Kerr interactions have been proposed by using the Purcell effect
\cite{KerrPE}, Rydberg atoms \cite{KerrRA}, interaction of a cavity mode
with atoms \cite{PlenioKerr} and nanomechanical resonators
\cite{KerrNMR1}.  Notice that the dynamics in a Kerr medium may be
accurately described in terms of the Wigner function in the phase-space
\cite{KerrWigMilburn}.
\par
In this paper we consider generic Gaussian states undergoing self-Kerr
interaction and investigate their use in estimation of displacement and
squeezing parameters. Indeed, displacement and squeezing are basic
Gaussian operations in continuous variable systems and represent
building blocks to manipulate Gaussian states for quantum information
processing.  Besides, they represent the ultimate description of
interferometric  interaction. As a consequence, their characterization,
{\em i.e} the optimal estimation of displacement and squeezing
parameters has been widely investigated
\cite{Hel74,Mil94,San95,Per01,Chi06,Gai09} by using different tools from
quantum estimation theory (QET)
\cite{QET1,QET2,QET3,HEL67,LQE1,LQE2,BC94,LQE}.
\par
Our main goal is to assess Kerr interaction and the resulting
non-Gaussianity (nonG) as a resource for parameter estimation, and to this aim
we consider two different situations with different physical
constraints. On the one hand we study schemes where we fix the overall
energy available to the probe, without posing any constraint on the
available Gaussian squeezing; this will be referred to as the fixed
energy case. On the other hand, we will analyze the more realistic case
where the amount of Gaussian squeezing is fixed, or even absent, and
refer to this case as the fixed squeezing case.  As we will see, at
fixed energy Gaussian squeezing still represents the optimal resource
for parameter estimation. On the other hand, when the amount of Gaussian
squeezing is fixed then Kerr interaction turns out to be useful to
improve estimation, especially when the probe states have a large number
of {\em non squeezing} photons, i.e large amplitude. In this case
precision obtained by Gaussian states is achieved or enhanced.
\par
The paper is structured as follows. In Sec. \ref{s:qet} 
we review few basic ingredients of local quantum estimation theory
and illustrate the content of the quantum Cramer-Rao bound. In Sec,
\ref{s:displa} we analyze the use of Kerr interaction to improve
estimation of the displacement amplitude, whereas in Sec.
\ref{s:squeeze} we focus on squeezing estimation. Sec. \ref{s:outro}
closes the paper with some concluding remarks.
\section{Quantum estimation theory}\label{s:qet}
Let us start by reviewing some basic concepts of local quantum
estimation theory: when a physical parameter is not directly accessible 
one has to resort to indirect measurements, i.e. , measuring an
observable somehow related to the quantity of interest and estimate its 
value from the experimental sample. Let us denote by $\lambda$ the 
quantity of interest, $X$ the measured observable, and $\chi=(x_1, \ldots, 
x_M)$ the observed sample. The {\em estimation problem} amounts to find an 
estimator, that is a map $\hat{\lambda}= \hat{\lambda}(\chi)$ from the 
set of the outcomes to the space of parameters. Classically, optimal 
estimators are those saturating the Cramer-Rao inequality 
$\textrm{Var}(\lambda)\geq [M F(\lambda)]^{-1}$, which bounds from 
below the variance $\textrm{Var}(\lambda)=E
[\hat{\lambda}^2]-E [\hat{\lambda}]^2 $ of any unbiased estimator
of the parameter $\lambda$. In the Cramer-Rao inequality, $M$ is the
number of measurements and $F(\lambda)$ is the so-called Fisher
Information (FI) $F(\lambda)= \int\! dx\, p(x|\lambda)\left[
\partial_\lambda \ln p(x|\lambda) \right ]^2 $ where $p(x|\lambda)$ is
the conditional probability of obtaining the value $x$ when the
parameter has the value $\lambda$.  The quantum analog of the
Cramer-Rao bound is obtained starting from the Born rule $p(x|\lambda)=
\Tr[\Pi_x \varrho_\lambda]$ where $\{\Pi_x\}$ is the probability
operator-valued measure (POVM) describing the measurement and
$\varrho_\lambda$ the density operator, labeled by the parameter of
interest. In order to evaluate the ultimate bounds to precision
one introduces the Symmetric Logarithmic Derivative (SLD) $L_\lambda$ as
the operator satisfying 
$2 \partial_\lambda\varrho_\lambda=  
L_\lambda \varrho_\lambda+ \varrho_\lambda L_\lambda$
and prove that the FI is upper bounded by the Quantum 
Fisher Information (QFI) \cite{BC94}
$ F(\lambda)\leq H(\lambda)\equiv\Tr[\varrho_\lambda L_\lambda^2].  $
In turn, the ultimate limit to precision is given by the quantum 
Cramer-Rao bound 
$\textrm{Var}(\lambda)\geq [M H(\lambda)]^{-1}$.
Let us consider the case where the parameter of interest is 
the shift imposed by a unitary evolution $U_\lambda= \exp(-i \lambda G)$ 
to a given initial pure state $|\psi_0\rangle$, $G$ being
the corresponding Hermitian generator.
The family of states we are dealing with 
is given by $|\psi_\lambda\rangle= U_\lambda |\psi_0\rangle$, 
and since for pure states $\varrho_\lambda^2=\varrho_\lambda$,
one has $\partial_\lambda 
\varrho_\lambda=\partial_\lambda \varrho_\lambda \varrho_\lambda+
\varrho_\lambda\partial_\lambda \varrho_\lambda$, i.e. 
\begin{align}
L_\lambda &= 2[\ketbra{\psi_\lambda}{\partial_\lambda
\psi_\lambda}+ \ketbra{\partial_\lambda\psi_\lambda}{\psi_\lambda}]
\notag \\ 
H(\lambda) &= 4 \left[ \braket{\partial_\lambda\psi_\lambda}{\partial_\lambda
\psi_\lambda} +
(\braket{\partial_\lambda\psi_\lambda}{\psi_\lambda})^2\right]
\notag\:.
\end{align}
After some algebra one sees that the QFI turns out to be proportional 
to the fluctuations of the generator on the probe state,
$H(\lambda)
= 4 \langle \psi_0|\Delta G^2 | \psi_0\rangle$, 
and thus it is independent on the value of $\lambda$. 
The above equation, together with the Cramer-Rao bound, expresses 
the ultimate quantum lower bound on the precision achievable by using 
a given probe $|\psi_0\rangle$ and any estimation procedure, i.e. without making
reference to any specific detection scheme. In the following we will
exploit the above tools to assess and compare the use of Gaussian states
and  Kerr modified Gaussian states in the estimation of displacement
and squeezing parameters.  More specifically, we evaluate the QFI as a
function of the involved parameters and analyze its behaviour in
different relevant regimes.
\section{Estimation of displacement}\label{s:displa}
Let us first consider the estimation of displacement, i.e. of the real 
parameter $\lambda \in \mathbbm{R}$ imposed by the unitary $U_\lambda =
\exp \{ -i \lambda G_d \}$, $G_d=a^{\dag} + a$ being the corresponding
generator. For a generic pure Gaussian probe, i.e. a 
displaced squeezed state of the form $\ket{\alpha,r} =D(\alpha)S(r)|0\rangle$ 
(with $\alpha= |\alpha|e^{i\phi}$ and $r>0$) where $D(\alpha) = \exp \{ 
\alpha a^{\dag} - \bar\alpha a ) \}$  and $S(r) = \exp\{\frac{r}{2}(a^{{\dag} 2}-a^2)\}$, 
the QFI, i.e. the fluctuations of the generator, may be evaluated by 
normal ordering for creation and annihilation operators \cite{Marian}. 
One obtains $H^{(d)}= 4 +
8 N \beta + 8\sqrt{ N\beta (1+N\beta)}$, where $N=\sinh^2 r +
|\alpha|^2$ is the number of photons of the probe state and where
$\beta=\sinh^2r /N$ is the corresponding squeezing fraction ($0 \le
\beta \le 1$ ). As expected for a unitary family the QFI does not depend
on the value of the parameter. Besides, the QFI depends only on the squeezing energy
$N_{sq}=\beta N$, and thus increasing the amplitude energy
$N_{\alpha}=|\alpha|^2$, does not lead to any enhancement of precision.
Therefore, at fixed energy, the maximum QFI $H_{S}^{(d)}= 4 + 8 N +
8\sqrt{ N (1+N)}$ is achieved for $\beta=1$, \emph{i.e.} for squeezed
vacuum. In the opposite limit ($\beta=0$),
\emph{i.e.} for coherent states, the QFI is constant:
$H_{C}^{(d)}=4$. 
Let us consider now a generic Gaussian state that undergoes Kerr interaction
$|\alpha , r ,\gamma\rangle = U_\gamma D(\alpha) S(r) |0\rangle$
where $U_\gamma = \exp ( -i \gamma (a^{\dag} a)^2 )$.
The QFI for this class of states can be evaluated numerically 
upon varying the parameters $\gamma$, $|\alpha|$, $\phi$, and $r$. 
We found that at fixed energy, the optimal 
probe state is still the squeezed vacuum state. The optimal 
QFI is a monotonous decreasing function of $\gamma$ and the Kerr 
dynamics does not improve estimation precision. In other words, at fixed energy, 
squeezed vacuum state is the best probe not only among the class of Gaussian
states, but also maximizing the QFI over the wider class of states Kerr perturbed 
Gaussian states. 
\par
\begin{figure}[h!]
\includegraphics[width=0.8\columnwidth]{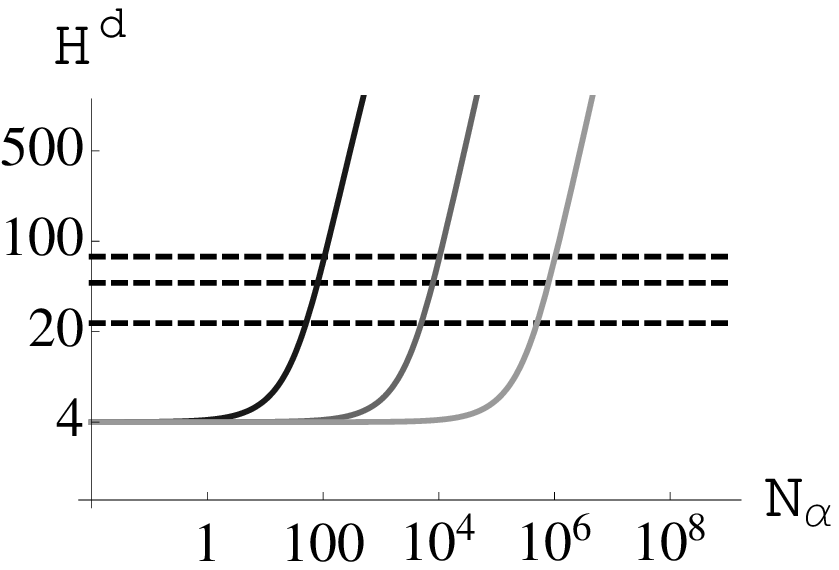}
\includegraphics[width=0.8\columnwidth]{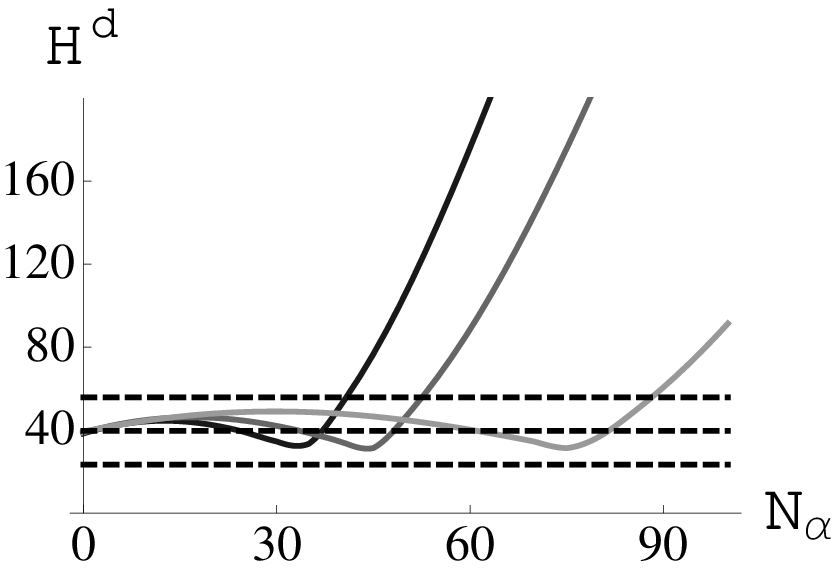}
\caption{\label{f:DispFixedSq} 
\underline{Top}: QFI $H_{\gamma}^{(d)}$ for displacement estimation by Kerr
modified coherent states (solid lines) as a function of the number of
photons $N_{\alpha}$ and for different values of $\gamma$. From darker to 
lighter gray: $\gamma= \{ 10^{-2},10^{-4},10^{-6} \}$. 
Dashed lines refer to QFI $H_{G}^{(d)}$ of squeezed
vacuum states for different values of squeezing photons. From bottom to
top: $N_{sq}=1,2,3$. \underline{Bottom}: QFI $H_{\gamma}^{(d)}$ for Kerr modified
displaced squeezed states, $N_{sq}=2$, for different values of $\gamma$.
From darker to lighter gray: $\gamma= \{ 0.01, 0.008, 0.005 \}$. 
Dashed lines denote QFI
$H_{G}^{(d)}$ of squeezed vacuum states for different values of
squeezing photons. From bottom to top: $N_{sq}=1,2,3$.}
\end{figure}

Let us now address estimation of displacement in the more realistic 
configuration, where the amount of Gaussian squeezing is fixed or absent.
For Kerr modified coherent states $|\alpha,\gamma\rangle$, 
QFI can be evaluated analytically at fixed energy $N=|\alpha|^2$ and $\gamma$,
arriving at
\begin{align}
H^{(d)}=& \: 4 + 8 N e^{-4 N \sin^2 \gamma} \Big\{
e^{4 N \sin^2 \gamma} -1 \notag \\  &+ \cos [2(\gamma-\phi + N
\sin 2\gamma)] \notag \\
&-e^{-4 N \cos 2\gamma \sin^2 \gamma}\cos[4 \gamma - 2 \phi + N
\sin 4 \gamma]
\Big\}
\end{align}
and then optimized numerically over the coherent phase $\phi$. 
The results are reported
in Fig. \ref{f:DispFixedSq} (top plot)as a function of the number of photons 
$|\alpha|^2$ and for different values of $\gamma$. 
The QFI increases with $|\alpha|^2$ and $\gamma$ and the precision achievable 
with current technology squeezing, say $N_{sq}\lesssim 2$, may be
attained and surpassed for realistic values of the Kerr coupling $\gamma$ 
and large enough signal amplitude, say $\gamma |\alpha|^2\lesssim 1$. 
Better performances may be obtained by considering Kerr modified 
squeezed states $|\alpha , r ,\gamma\rangle$ with fixed squeezing $r$ 
and large amplitude $|\alpha|\gg 1$. The QFI for this case, as evaluated 
numerically and optimized over the amplitude phase $\phi$ is
reported in Fig. \ref{f:DispFixedSq} (bottom plot). We observe that, after 
a regime where QFI oscillates around the value obtained for 
vanishing $\gamma$, then it increases monotonically with $|\alpha|^2$ and 
exceed the corresponding Gaussian QFI for large enough values
of $|\alpha|^2$ and/or $\gamma$. Due to numerical limitations,
we have considered $|\alpha|^2\leq 100$, and thus we have seen 
enhancement of precision only for the largest values of $\gamma$. 
We expect analog performances by considering smaller values of 
$\gamma$ and  larger numbers of photons.
\section{Estimation of squeezing}\label{s:squeeze}
Let us now consider estimation of squeezing, that is the estimation of
the real parameter $z \in \mathbbm{R}$ imposed by the unitary evolution
$U_z = \exp \{ -i z G_s \}$ with generator $G_s=\frac12(a^{\dag
2}+a^2)$.  Given a generic single-mode Gaussian state $|\alpha,r\rangle$, 
the QFI for squeezing estimation has been
evaluated by using the normal
ordering for creation and annihilation operators \cite{Marian}. 
The maximum is $H^{(s)}_G=8N^2+8N+2$ and is again achieved using 
squeezed vacuum probe \cite{Gai09}. In order to investigate the 
effect of Kerr interaction we
consider Kerr modified Gaussian states $|\alpha , r ,\gamma\rangle$.  
At fixed energy QFI has been evaluated and 
optimized numerically against the squeezing fraction $\beta$ and phase 
$\phi$. In this case, the optimal squeezing fraction decreases monotonically 
with both $\gamma$ and the total number of photons $N$ and the maximized QFI 
is a decreasing function of $\gamma$, that is Kerr interaction does not 
improve, actually degrades, the estimation precision achievable with squeezed 
vacuum probe. 
\begin{figure}[h!]
\includegraphics[width=0.75\columnwidth]{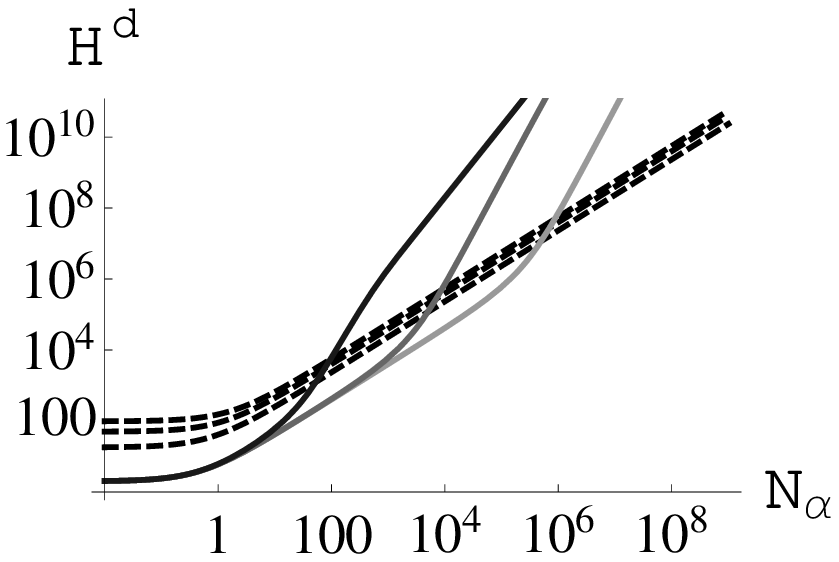}
\includegraphics[width=0.80\columnwidth]{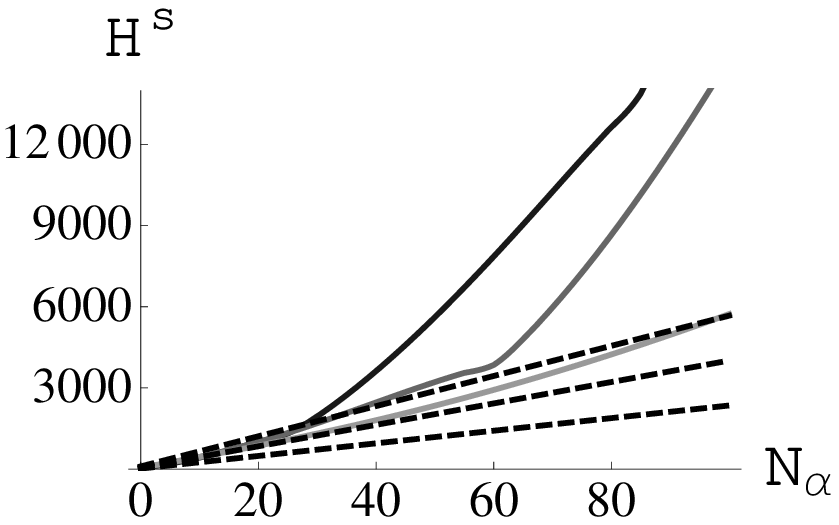}
\caption{\label{f:SqueezedFixedSq}
\underline{Top}: QFI $H_{\gamma}^{(s)}$ for squeezing
estimation by Kerr modified coherent probes (solid lines) as a function
of the number of photons $N_{\alpha}$ and for different values of
$\gamma$. 
From darker to lighter gray: $\gamma= \{ 10^{-2},10^{-4},10^{-6} \}$. 
Dashed lines refer to the QFI
$H_{G}^{(s)}$ for displaced squeezed probes and different values of
squeezing photons. From bottom to top: $N_{sq}=1,2,3$.
\underline{Bottom}: QFI $H_{\gamma}^{(s)}$ for Kerr modified displaced
squeezed states (solid lines) with $N_{sq}=2$ squeezing photons, as a
function of field amplitude photons $N_{\alpha}=|\alpha|^2$ and for
different values of $\gamma$.
From darker to lighter gray: $\gamma= \{ 0.01 ,0.005, 0.001 \}$. 
Dashed lines refer to the QFI
$H_{G}^{(s)}$ for displaced squeezed vacuum states and different values
of squeezing photons. From bottom to top: $N_{sq}=1,2,3$.}
\end{figure}
\par
Let us now consider situations where squeezing is not available, or its
amount is fixed, and where the field amplitude may be increased at will.
The QFI for probe states of the form $|\alpha,\gamma\rangle=U_{\gamma}
D(\alpha)|0\rangle$ can be evaluated analytically as 
\begin{align}
H^{(s)}=& \: 2 + 2 N \Big\{ 2 + N \notag \\ &- N e^{-4 N \sin^2 \gamma} 
(1+ \cos [2(4\gamma-2 \phi + N
\sin42\gamma)] )
\notag \\  &+ N e^{-N (1- \cos 8\gamma) }
\cos[16 \gamma - 4 \phi + N \sin 8 \gamma]
\Big\}\:,
\end{align}
and then maximized
numerically over the amplitude phase $\phi$. In Fig.
\ref{f:SqueezedFixedSq} we report the optimized QFI together with the
QFI of displaced squeezed vacuum states with $N_{sq}\leq 3$ and the same
value of $|\alpha|^2$. Results indicate that upon using coherent states
with large amplitude we may achieve and improve the precision of
squeezed vacuum states already for small, realistic, values of the Kerr
coupling $\gamma$.  When the amount of Gaussian squeezing is
nonzero but fixed we can combine the effects of squeezing and Kerr
interaction by considering Kerr modified displaced squeezed states with
a large number of amplitude photons ($|\alpha|^2\gg1$). As it is
apparent from Fig. \ref{f:SqueezedFixedSq} the QFI increases with
$|\alpha|^2$ and overtake quite rapidly the values of QFI of the
corresponding Gaussian state.
\section{NonGaussianity as an overall indicator of precision enhancement}
As pointed out in the introduction, Kerr interaction induces a 
nonGaussian operation. A question thus arises on whether there is 
a connection between the amount of nonG of the probe 
and the precision of estimation. In other words, whether or not nonG 
may be used as an overall indicator of precision enhancement due to
Kerr interaction. The answer to this question amounts to investigate the
behavior of the QFI as a function of a nonG measure. Different measures
of nonG for a quantum state have been recently introduced
\cite{nonGHS,nonGRE,nonGSimon} and here we consider the entropic measure
\cite{nonGRE} $\delta[\varrho]=S(\tau) - S(\varrho)$ where $S(\varrho)$
is the Von Neumann entropy of the state $\varrho$, and $\tau$ denotes
the Gaussian states with the same covariance matrix of the state
$\varrho$ under investigation. Since both, nonG for Kerr modified
coherent states  and the corresponding QFI are increasing functions of
the number of photons, we consider a \emph{normalized} nonG measure
$\delta_R[\varrho]= \delta[\varrho]/ \delta_m (N_\alpha)$, obtained as
the ratio between $\delta[\varrho]$ and the maximum nonG $\delta_m
(N_\alpha)$ achievable with the same number of photons. This is in order
to discern the real contribution of nonG to the improvement of
estimation from that coming from energy scaling. 
\begin{figure}[h!]
\includegraphics[width=0.80\columnwidth]{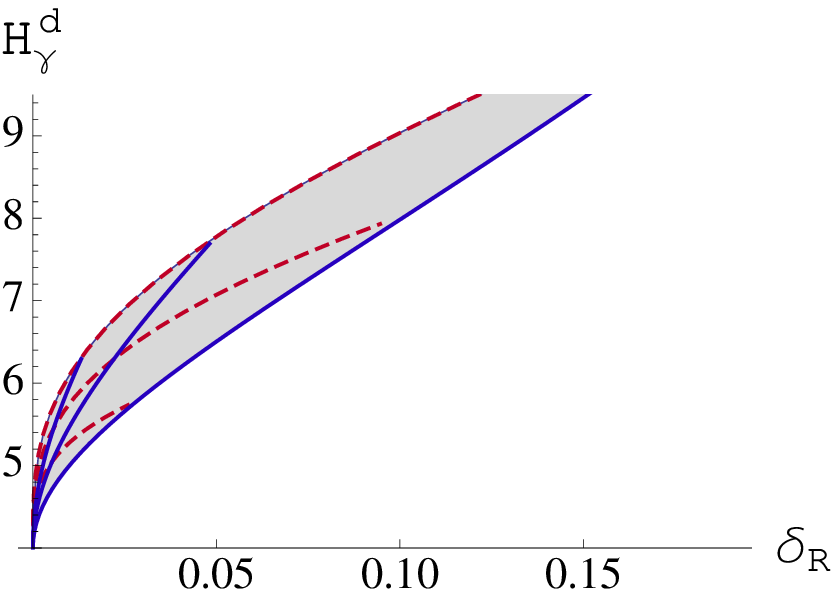}
\includegraphics[width=0.80\columnwidth]{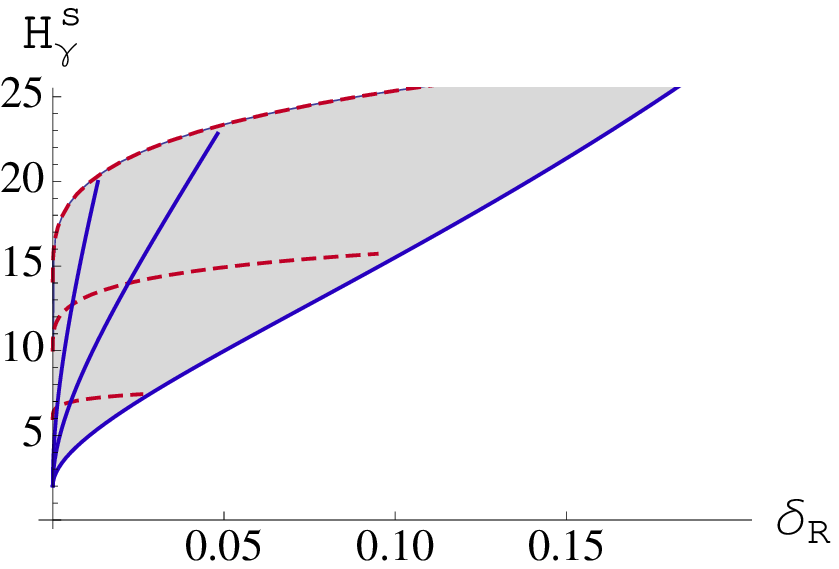}
\caption{
\label{f:nonGvsH}
(Color online) \underline{Top}: QFI $H_{\gamma}^{(d)}$ for displacement
estimation by Kerr modified coherent states as a function of the
normalized non-Gaussianity $\delta_R$.  The solid blue lines refer to
the case of fixed Kerr coupling $\gamma$ and varying number of photons
$0<N_{\alpha}<3 $; from top to bottom we have $\gamma= 0.04, 0.06,
0.10$.  The dashed red lines are for fixed number of photons and varying
Kerr coupling $0<\gamma<0.1 $; from top to bottom we have
$N_\alpha=3,2,1$. The gray area denotes the allowed values of both QFI
and nonG for the considered values of the amplitude and the coupling.
\underline{Bottom}: The same as in the left panel for the QFI
$H_{\gamma}^{(s)}$ for squeezing estimation.}
\end{figure}
\par
In Fig.
\ref{f:nonGvsH} we report the QFI for both displacement and squeezing
estimation by Kerr modified coherent states, as a function of the
normalized nonG for fixed Kerr constant $\gamma$ (varying the number of
photons $N_\alpha$) and for fixed number of photons $N_\alpha$ (varying
$\gamma$). As it is apparent from the plots, QFI is not a fully monotone
function of nonG: the allowed region for the values of parameters we
have considered is the gray area and one may find two states such that
$\delta_R[\varrho_1]>\delta_R[\varrho_2]$ and $H[\varrho_1] <
H[\varrho_2]$. On the other hand, if we fix one of the two parameters
($\gamma$ or $N_\alpha$) and vary the other one, we observe a monotonous
behavior. In other words nonG is quantitatively related to the 
increase in QFI and thus represents a good indicator to assess Kerr
interaction in quantum estimation.
\par
One may also ask whether the enhancement in precision obtained 
with Kerr interaction may be ascribed to the squeezing effect occurring
in Kerr evolution  at small time/nonlinearity and/or small number of 
photons. For coherent input, this is definitely not the case, as it can 
easily checked by noting that improvement in precision occurs for 
$\gamma |\alpha |^2 \lesssim 1$ i.e. when the state is no longer squeezed 
(see  \cite{KerrWigMilburn} for a phase space picture of Kerr evolution 
for input coherent states). Moreover, when we consider a Kerr perturbed 
squeezed state as input, the QFI for displacement estimation  is not monotone 
in the region where one may expect a further squeezing  effect or at least 
that the initial squeezing is conserved. Also in this case, enhancement
in precision is observed for increasing amplitude photons or Kerr 
nonlinearity, when the quantum state is no longer squeezed. 
At the same time, improvement is not due to the evolution towards 
cat states, since they are achieved by Kerr interaction only for very high 
nonlinearities and they present a different scaling in precision \cite{prep}.
The most intuitive picture one may draw is that the involved structure of
the Wigner function leads to its spread over the phase-space and 
consequently to a smaller overlap when displaced (squeezed). One should also 
notice that, since the phase of the 
coherent input signal is optimized for each pair of values of the coupling 
and the amplitude, a simple picture in terms 
of Wigner evolution may be even confusing rather than help intuition. 
For these reasons we consider nonG as a suitable quantity to
summarize the improvement in the estimation precision for Kerr perturbed
Gaussian states.

\section{Conclusions} \label{s:outro}
In conclusion, we have addressed the use of Kerr interaction to improve
estimation of displacement and squeezing parameters and analyzed in
details the behavior of the quantum Fisher information as a function
of probe and interaction parameters. We found that
at fixed energy, with no constraint on the available Gaussian squeezing,
Kerr dynamics is not useful and performances of Gaussian states are
superior.  On the other hand, in the more realistic case where the
amount of Gaussian squeezing is fixed, or absent, then Kerr
interaction improves estimation, especially for probe states with
large amplitude.
\par
It should be noticed that Gaussian squeezing in $\chi^{(2)}$ media is 
obtained by parametric processes and the amount of squeezing 
linearly increases with the pump intensity. On the other hand, 
in $\chi^{(3)}$ media, the energy needed to obtain significant nonlinear
effects is provided by the signal itself. Overall, our results indicate 
that precision achievable with current technology Gaussian squeezing 
may be attained  and surpassed for realistic values of the Kerr coupling 
and large enough signal amplitude.
We also found that precision improvement is 
quantitatively related with the amount of nonGaussianity induced by Kerr
interaction, and thus conclude that Kerr nonGaussianity is a resource,
achievable with current technology, for high-precision measurements.
We foresee a possible widespread use as a characterization tools in 
emerging quantum technologies like quantum communication and metrology.
\section*{Acknowledgements}
We thank Stefano Olivares for several discussions. This work 
has been partially supported by the CNR-CNISM
convention.

\end{document}